\documentclass[10pt,letterpaper,twocolumn]{article} %% two column, final layout

\usepackage{ol2}
\usepackage[draft]{hyperref}
\usepackage{amsmath}

\begin{document}

\twocolumn[ %% activate for two-column option

\title{Rabi oscillations of bound states in the continuum}

%% For REVTeX it is possible to automate superscript and e-mail callouts with the superscriptaddress option; see REVTeX4 documentation.

\author{Stefano Longhi}
\address{Dipartimento di Fisica, Politecnico di Milano and Istituto di Fotonica e Nanotecnologie del Consiglio Nazionale delle Ricerche, Piazza L. da Vinci 32, I-20133 Milano, Italy (stefano.longhi@polimi.it)}
\address{IFISC (UIB-CSIC), Instituto de Fisica Interdisciplinar y Sistemas Complejos, E-07122 Palma de Mallorca, Spain}

\begin{abstract}
Photonic bound states in the continuum (BICs) are special localized and non-decaying states of a photonic system with a frequency embedded into the spectrum of scattered states. The simplest photonic structure displaying a single BIC is provided by two waveguides side-coupled to a common waveguide lattice, where the BIC is protected by symmetry. Here we consider such a simple photonic structure and show that, breaking mirror symmetry and allowing for non-nearest neighbor couplings, a doublet of quasi-BIC states can be sustained, enabling weakly-damped embedded Rabi oscillations of photons between the waveguides. 
\end{abstract}

%\ocis{130.2790,  130.3120, 000.160)}
 ] %% activate for two-column option

{\em Introduction.} Interference associated with bound states in the continuum (BICs) has attracted much attention in photonics over the past decade  \cite{r1,r2,r3,r4}, with many technological implications for nanophotonics, quantum optics, laser design and nonlinear optics (see, e.g. \cite{r1,r2,r3,r4,r5,r5a,r6,r6b,r7,r8,r9,r10,r11,r12,r13,r14,r15,r16,r17,r18,r19,r20,r21,r22,r23,r24,r25,r26,r27,r28,r29,r30,r31} and references therein). Since a pioneering work on destructive interference in the radiation loss of a laser device \cite{r5}, several mechanisms have been proposed and demonstrated for realizing BICs in photonics \cite{r1}, including symmetry-protected BICs \cite{r1,r7,r8,r14}, BICs via separability \cite{r1,r18}, Fano or Fabry-P\`erot BICs \cite{r5a,r6,r7,r12},  and BICs from inverse engineering \cite{r9,r11}. While some BICs (like those observed in photonic crystal slabs \cite{r10}) are protected topologically and cannot be removed except by large variations in the parameters of the system \cite{r15}, perturbations typically turn a BIC into a leaky resonance, i.e. a quasi-BIC. Two waveguides W$_1$ and W$_2$ side-coupled to a photonic lattice in the geometrical settings of Fig.1(a)  provide the simplest optical system that sustains a symmetry-protected BIC \cite{r1}, as demonstrated in early experiments on optical BICs \cite{r7,r8}. In such systems, the BIC becomes a leaky mode (quasi BIC) when the mirror symmetry is broken \cite{r1,r8}, for  example by bending the waveguides \cite{r7} or by introduction of a transverse refractive index gradient \cite{r8}. For a photonic system sustaining two bound states, Rabi flopping between such modes is possible when they  are indirectly coupled via a common continuum. While this phenomenon is known for bound states {\em outside} the continuum \cite{r6}, so far Rabi flopping between BICs remains unexplored. In this Letter we introduce the idea of Rabi flopping between BICs, observed when two degenerate BICs are indirectly coupled via a common continuum of states resulting in a doublet quasi-BIC state.\\
 \\
{\em Model and asymptotic analysis.} To demonstrate the possibility of coupling two BICs with negligible radiation into the continuum of states, we extend the geometrical setting of Fig.1(a) \cite{r7,r8}, breaking mirror symmetry and allowing for non-nearest neighbor (multiple) couplings of the lateral waveguides W$_1$ and W$_2$ with the lattice [Fig.1(b)]. Photon propagation along the spatial axis $z$ of the waveguide-array system is described by the tight-binding Hamiltonian (see for instance \cite{r32,r33})
 \begin{equation}
 \hat{H}=\hat{H}_s+\hat{H}_{b}+\hat{H}_{int}
 \end{equation}
 \begin{figure}[htb]
 \centerline{\includegraphics[width=8.7cm]{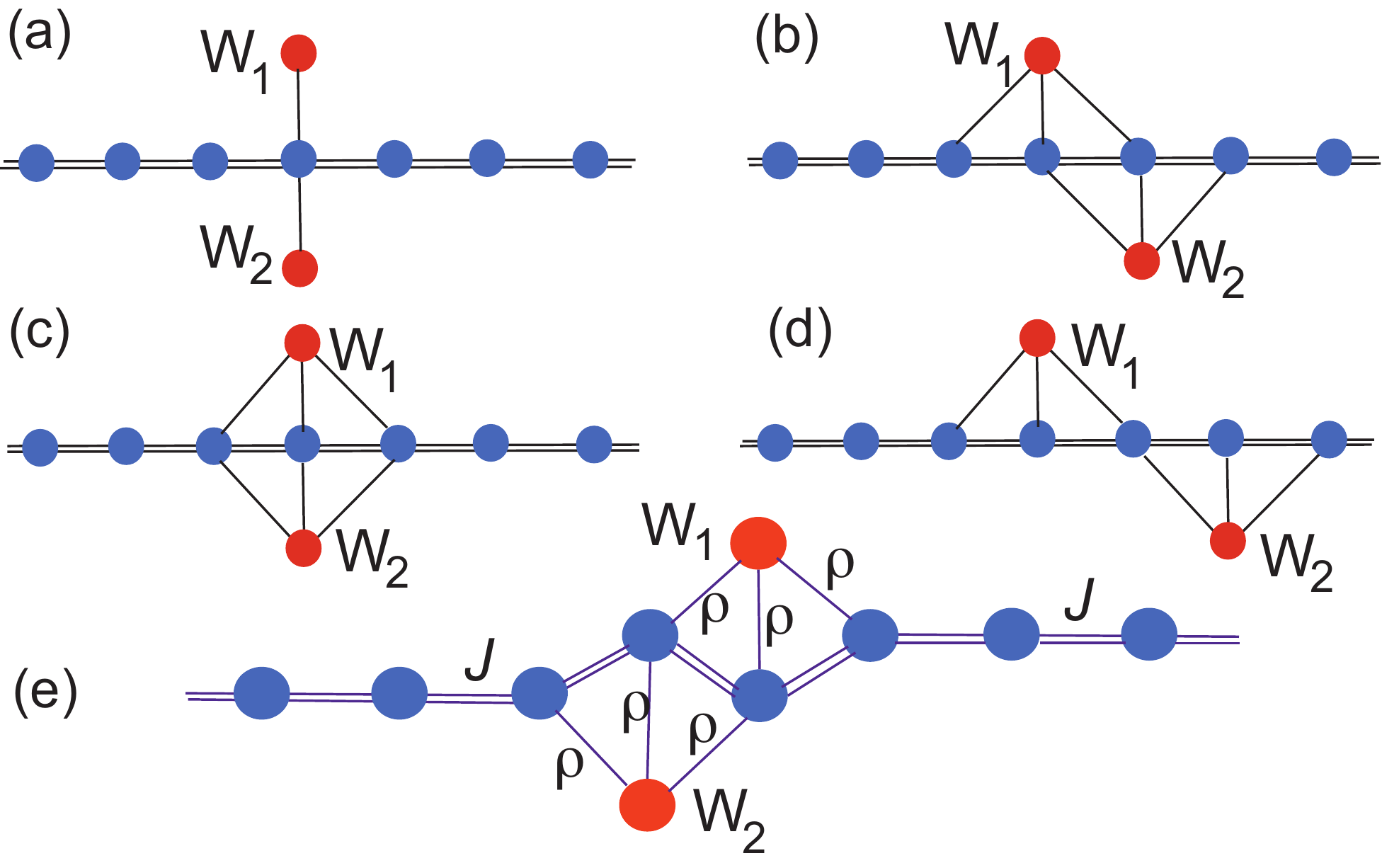}} \caption{ \small
 (Color online) (a) Two optical waveguides W$_1$ and W$_2$ weakly-coupled to a one-dimensional linear waveguide array under mirror symmetry with a single contact (coupling) point. The system sustains one symmetry-protected BIC. (b) Same as (a), but with three contact points and one-unit-cell spatial shift. The system can sustain two non-degenerate quasi-BICs. (c) Same as (a) but with three contact points. (d) Same as (b), but for a spatial shift of two unit cells. In (c) and (d) the system does not show Rabi flopping. (e) Practical design of a photonic lattice which allows equal coupling constants $\rho$ for the three contact points. The solid lines (bonds) in the panels indicate mode coupling of attached waveguides.}
 \end{figure} 
 where $\hat{H}_s= \omega_0 (\hat{a}_1^{\dag} \hat{a}_1+ \hat{a}_2^{\dag} \hat{a}_2)$ is the Hamiltonian of the photon field in waveguides W$_1$ and W$_2$, 
  $\hat{H}_b=  J \sum_n ( \hat{c}^{\dag}_n \hat{c}_{n+1}+ H.c.)$ is the tight-binding Hamiltonian of the photonic modes  of the lattice, and 
 \begin{equation}
 \hat{H}_{int}= \sum_{n} \left(  \rho_n \hat{a}_1^{\dag} \hat{c}_{n}+ \sigma_n \hat{a}_2^{\dag} \hat{c}_{n} + H.c.\right)
 \end{equation}
  is the Hamiltonian describing evanescent mode coupling of waveguides W$_1$  and W$_2$ with the lattice. In the above equations, $\hat{a}_1^{\dag}$ and $\hat{a}_2^{\dag}$ are the bosonic creation operators of photons in waveguide W$_1$ and W$_2$, respectively,  $\hat{c}^{\dag}_n$ is the bosonic creation operator of photons in the $n$-th waveguide of the lattice, satisfying the usual commutation relations $ [\hat{a}_n, \hat{a}_l^{\dag}]=\delta_{n,l}$, $ [\hat{a}_l, \hat{c}_n^{\dag}]=0$, $[ \hat{c}_l, \hat{c}^{\dag}_n]= \delta_{n,l}$, etc.; $J$ is the coupling constant between adjacent waveguides in the lattice; $\omega_{0}$ is the propagation constant shift of the modes in waveguides W$_{1,2}$ from the propagation constant of the lattice waveguides; $\rho_n$ is the coupling constant between waveguide W$_1$ and the $n$-th lattice waveguide; and   $\sigma_n$ is the coupling constant between waveguide W$_2$ and the $n$-th lattice waveguide.  We assume $| \omega_0|<2J$, so that rather generally in the weak coupling limit photons can escape from the waveguides W$_{1,2}$ into the lattice. To describe photon dynamics in the system, it is worth switching from Wannier to Bloch basis representation for the photon modes in the lattice \cite{r33}, that diagonalizes $\hat{H}_b$. After introduction of the bosonic operators in Bloch basis
 \begin{equation}
 \hat{C}(k)  \equiv  \frac{1}{ \sqrt {2 \pi}}  \sum_n \hat{c}_n \exp(-ikn) 
 \end{equation}
 where $-\pi \leq k < \pi$ is the Bloch wave number, the full Hamiltonian of the photon field is given by Eq.(1) with
 \begin{eqnarray}
\hat{H}_s & = & \omega_0 (\hat{a}_1^{\dag} \hat{a}_1+ \hat{a}_2^{\dag} \hat{a}_2) \nonumber \\
\hat{H}_b & = & \int_{-\pi}^{\pi} dk \; \omega(k) \hat{C}^{\dag}(k) \hat{C}(k) \\
\hat{H}_{int} & = & \int_{-\pi}^{\pi} dk \left\{ G_1(k) \hat{a}_1^{\dag} \hat{C}(k)+G_2(k) \hat{a}_2^{\dag} \hat{C}(k) +H.c.  \right\} \nonumber
\end{eqnarray}
where 
\begin{equation}
\omega(k) \equiv 2 J \cos k
\end{equation}
is the dispersion relation of the tight-binding energy band and $G(k), F(k)$ are the spectral coupling functions, given by
\begin{eqnarray}
G_1(k) & = & \frac{1}{\sqrt{2 \pi}} \sum_n \rho_n \exp(i kn) \nonumber \\
G_2(k) & = & \frac{1}{\sqrt{2 \pi}} \sum_n \sigma_n \exp(i kn).
\end{eqnarray}
Let us assume that the waveguide-array system is initially excited by a single photon or, likewise, by a classical state of light \cite{r6b,r32}. In this case we can restrict the analysis to the single excitation sector of Fock space by letting 
\begin{eqnarray}
 | \psi(z) \rangle  =  \exp(-i \omega_0 z) \times \\ 
  \left\{ a_1(z) \hat{a}_1^{\dag} | 0 \rangle + a_2(z) \hat{a}_2^{\dag} | 0 \rangle+ \int_{-\pi}^{\pi} dk c(k,z) \hat{C}^{\dag}(k) |0 \rangle \right\}  \nonumber 
\end{eqnarray}
for the state vector of the photon field, 
where the amplitude probabilities $a_1(z)$, $a_2(z)$ and $c(k,z)$ satisfy the (classical) coupled-mode equations
\begin{eqnarray}
i \frac{ da_{1,2}}{dz} & = & \int_{-\pi}^{\pi} dk \; G_{1,2}(k) c(k,z)  \\
%i \frac{ da_2}{dz} & = & \int_{-\pi}^{\pi} dk \; G_2 (k) c(k,z) \\
i \frac{\partial c}{\partial z} & = & \left\{  \omega(k)-\omega_0 \right\} c(k,z)+G_1^*(k)a_1+G_2^*(k)a_2. \nonumber
\end{eqnarray}
We assume that the photon is injected with some amplitude probabilities $a_1(0)$, $a_2(0)$ in either waveguides W$_1$ or W$_2$, i.e. we assume $c(k,0)=0$. The {\em exact} evolution of the amplitudes $a_1(z)$, $a_2(z)$ can be formally calculated from Eqs.(8) by standard Laplace transform methods \cite{r33,r34,r35}, leading to coupled differential-delayed equations \cite{r33,r35b}. A simplified analysis, which highlights the emergence of BIC states in the waveguide-array system, is obtained in the  
% One obtains
%\begin{equation}
%a_{1,2}(z)=\frac{1}{2 \pi i } \int_{-i \infty+o^+}^{i \infty+o^+} ds \; A_{1,2}(s) {\exp(sz)}
%\end{equation}
%where the Laplace transforms $A_1(s)$ and $A_2(s)$ read
%\begin{eqnarray}
%A_{1}(s) & = &  \frac{ia_1(0) (\Sigma_{2,2}-is)-i a_{2}(0) \Sigma_{1,2}} {\Sigma_{1,2} \Sigma_{2,1} -(\Sigma_{1,1}-is) (\Sigma_{2,2}-is) } \\
%A_{2}(s) & = &  \frac{ia_2(0) (\Sigma_{1,1}-is)-i a_{1}(0) \Sigma_{2,1}} {\Sigma_{1,2} \Sigma_{2,1} -(\Sigma_{1,1}-is) (\Sigma_{2,2}-is) }. 
%\end{eqnarray}
%In the above equations, the self-energy matrix $\Sigma_{l,n}$ is defined by
%\begin{eqnarray}
%\Sigma_{l,n}(s)=\int dk \frac{G_l(k) G_n^*(k)}{is-\omega(k)+\omega_0}
%\end{eqnarray}
%($l,n=1,2$). Note that, since the coupling constants $\rho_n$, $\sigma_n$ are real positive parameters and $\omega(-k)=\omega(k)$, it readily follows that the self-energy matrix is symmetric, i.e. $\Sigma_{1,2}(s)=\Sigma_{2,1}(s)$.
  weak-coupling approximation $\rho_n, \sigma_n \ll J$ and neglecting retardation (delay) effects \cite{r33}. Under such assumptions
 the amplitude probabilities $a_1(z)$ and $a_2(z)$ satisfy the approximate coupled-mode equations (see Supplemental)
\begin{equation}
i \frac{da_1}{dz}=\Sigma_{1,1}a_1+\Sigma_{1,2}a_2 \; , \; i \frac{da_2}{dz}=\Sigma_{2,1}a_1+\Sigma_{2,2}a_2
\end{equation}
where 
\begin{eqnarray}
\Sigma_{1,1} = -(i/v_g) \sum_{n,l} \rho_n \rho_l \exp(-ik_0|n-l|) \nonumber \\
\Sigma_{1,2}  = \Sigma_{2,1}= - (i/v_g) \sum_{n,l} \rho_n \sigma_l \exp(-ik_0|n-l|) \\
\Sigma_{2,2}  =  -({ i}/{v_g}) \sum_{n,l} \sigma_n \sigma_l \exp(-ik_0|n-l|) \nonumber
\end{eqnarray}
are the elements of the self-energy matrix, $k_0>0$ is the Bloch wave number defined by the resonance condition
\begin{equation}
2 J \cos k_0= \omega_0 \; ,
\end{equation}
and  $v_g=2 J \sin k_0$ is the group velocity of the radiating field.\\
\\
{\em Bound states in the continuum and Rabi flopping.}
Indicating by $\lambda_{1,2}$ the two eigenvalues of the self-energy matrix 
%\[
%\lambda_{1,2}= \frac{\Sigma_{1,1}+\Sigma_{2,2}}{2} \pm \sqrt{ \left( \frac{\Sigma_{1,1}+\Sigma_{2,2}}{2} \right)^2  +\Sigma_{1,2} \Sigma_{2,1}-\Sigma_{1,1}\Sigma_{2,2}}
%\]
 with ${\rm Im}(\lambda_1) \leq {\rm Im}(\lambda_2) \leq 0$, in the weak-coupling approximation and neglecting retardation effects the system sustains zero, one or two BICs depending on whether ${\rm Im}(\lambda_{1,2})<0$ (zero BIC), ${\rm Im}(\lambda_{2})=0$ and ${\rm Im}(\lambda_{1})<0$ (one BIC),  ${\rm Im}(\lambda_{1})={\rm Im}(\lambda_{2})=0$ (two BICs). 
When $\Sigma_{1,1} \Sigma_{2,2}=\Sigma_{1,2} \Sigma_{2,1}$, the system sustains one BIC with zero eigenvalue $\lambda_2=0$. This is the case of the geometrical setting of Fig.1(a), where $\Sigma_{1,1}=\Sigma_{2,2}=\Sigma_{1,2}=\Sigma_{2,1}$. The BIC is an anti-symmetric mode and protected by mirror symmetry. However,  in such a geometry one cannot realize Rabi flopping between the two indirectly-coupled waveguide W$_1$ and W$_2$, since the other (symmetric) mode of Eq.(9) is leaky. Interestingly, when non-nearest neighbor couplings are non-negligible and mirror symmetry is broken [Fig.1(b)], under certain conditions the system can sustain {\it two} BICs, which are indirectly coupled via the lattice,  thus allowing for undamped Rabi flopping between waveguide W$_1$ and W$_2$ mediated by the continuum of modes in the array. Rather generally, a BIC doublet is found whenever all the elements of the self-energy matrix are {\rm real} numbers and $\Sigma_{1,2} \neq 0$. The condition ${\rm Im}(\Sigma_{1,1})={\rm Im}(\Sigma_{2,2})=0$ means that each waveguide side-coupled to the lattice, i.e. either W$_1$ or W$_2$ in the absence of the other waveguide, does not radiate into the array. This is clearly impossible when the waveguide is side-coupled to a  single-site of the lattice, but not when we allow for multiple couplings (contact points), as in Fig.1(b). For example, let us assume that the waveguide W$_1$ is evanescently coupled to the three guides $n=-1,0,1$ (three contact points) of the lattice with the coupling constants  $\rho_0= \rho$, $\rho_1=\rho_{-1}=h \rho$, and $\rho_n=0$ for $n \neq 0, \pm1$, where $h$ is a dimensionless parameter that measures the relative strength of non-nearest-neighbor coupling. The corresponding expressions of the self-energy matrix elements [Eq.(10)] read 
\begin{eqnarray}
\Sigma_{1,1} & = & \Sigma_{2,2}=-\frac{i \rho^2}{2 J \sin k_0} \left\{  1+2h^2 \right. \nonumber \\
& + & \left. 4h \exp(-ik_0)+2h^2 \exp(-2ik_0) \right\} \nonumber
\end{eqnarray}
\begin{eqnarray}
\Sigma_{1,2} & = & \Sigma_{2,1}=-\frac{i \rho^2}{2 J \sin k_0}   \left\{ 2h+(1+3h^2) \exp(-ik_0)  \right. \nonumber \\
 &+ &  \left. 2h \exp(-2ik_0) +h^2 \exp(-3ik_0) \right\} \nonumber
\end{eqnarray}
From the above relations, it readily follows that all the self-energy matrix elements are real provided that
\begin{equation}
2 \cos k_0=-1/h,
\end{equation}
 i.e. $\omega_0=2J \cos k_0=-J/h$, which necessarily requires $h>1/2$. In this case the two BICs correspond to states with definite symmetry, i.e. $a_1=a_2$ for the symmetric BIC mode and $a_2=-a_1$ for the antisymmetric BIC mode. Their frequency splitting is the frequency of Rabi flopping, given by 
\begin{equation}
\Omega_R  =  2 | \Sigma_{1,2}| = {2 h^2 \rho^2}/{J}.
\end{equation}
We note that, if the system is initially excited by a non-classical state of light, for example in a biphoton state, the frequency of Rabi flopping is doubled \cite{r6b,r32}. Also, we note that, if the second waveguide W$_2$ were connected to the array without a spatial shift, i.e. without breaking the mirror symmetry [Fig.1(c)], the system would sustain two degenerate BICs because $\Sigma_{1,2}=0$, so that Rabi flopping would be prevented since $\Omega_R$ diverges. A similar behavior would be observed for a spatial shift larger than one unit cell [as in Fig.1(d)].\\
The previous asymptotic analysis is strictly valid in the weak coupling regime and neglects retardation effect. When such effects are considered, 
an exact scattering analysis of the waveguide-array scheme of Fig.1(b) shows that the two BICs become resonant states with a long lifetime, i.e. quasi-BICs. In the special case $h=1$ the antisymmetric BIC survives while the symmetric BIC becomes a quasi BIC; technical details are given in the Supplemental. Since the damping rate of the quasi BICs is much smaller than their frequency separation, weakly-damped Rabi oscillations are observed. The same scenario is expected to arise when the propagation constant shift $\omega_0$ is not exactly tuned to the value $-J/h$, due to e.g. practical imperfections, or when higher-order coupling terms (i.e. more than three contact points) are considered.\\
\\
{\em Numerical results and array design.} We checked the existence of weakly-damped Rabi flopping between quasi BICs by numerically solving the exact coupled-mode equations (8), that describe light transport in the lattice, with the initial condition $a_1(0)=1$, $a_2(0)=0$, $c(k,0)=0$, corresponding to photon excitation of waveguide W$_1$ at the input plane. In the numerical simulations, we assumed $h=1$, i.e. $\rho_n=\rho$ for $n=0, \pm1$, $\rho_n=0$ otherwise. The asymptotic analysis predicts a BIC doublet at $\omega_0=-J$, while the exact analysis shows that the symmetric BIC becomes actually a quasi-BIC with an imaginary part of the complex frequency (damping) much smaller than the Rabi frequency $\Omega_R$. In a typical waveguide lattice with circular symmetry of waveguide modes, the coupling constant between two closely-spaced waveguides is a nearly exponential function of waveguide spacing, so that the condition $h=1$ can be practically realized in the geometrical setting schematically shown in Fig.1(e), where waveguide W$_1$ (and likewise W$_2$) is equally-distant from three waveguides of the array, while the spacing between the waveguide centers in the array is uniform. Figure 2 shows typical numerical results corresponding to a weak-coupling [$\rho/J=0.2$, Fig.2(a)] and moderate-coupling [$\rho/J=0.4$, Fig.2(b)] regimes. The three curves in each panel correspond to different values of the propagation constant shift $\omega_0$, namely the resonant shift predicted by the asymptotic analysis $\omega_0=-J$ (curves 1), and other two values that deviate from $\omega_0=-J$  by 10$\%$ (curves 2 and 3, corresponding to $\omega_0/J=0.9$ and $\omega_0/J=-1.1$, respectively). In the weak-coupling regime and at exact resonance $\omega_0=-J$, the system clearly displays nearly undamped Rabi oscillations, with a frequency in good agreement with the theoretical prediction given by Eq.(13) ($\sim 0.077 J$ versus the theoretical value $0.08 J$); the damping rate $\gamma$ of the quasi BIC is in fact extremely small ($\Omega_R/ \gamma \sim 1228$) , as shown in the Supplemental. When the propagation constant shift $\omega_0$ is varied by the resonance value by about 10$\%$ [curves 2 and 3 in Fig.2(a)], the Rabi oscillations become weakly damped. Note that the revival probability $|a_1(z)|^2$ does not exactly return to the initial value $1$, even for exact resonance $\omega_0=-J$ where damping of Rabi flopping is not visible. This is mainly due to the fact that the initial condition has some non-vanishing overlapping (albeit small, of order $ \sim \rho/J$) with radiation modes. 
 In the moderate coupling regime $\rho=0.4J$, shown in Fig.2(b), radiation into the array is clearly observed in the initial stage of the dynamics and Rabi oscillations are weakly-damped, even at resonance. Also, superimposed high-frequency modulations are observed, which are due the back-flow of excitation from the array to the guides which is a typical feature of the moderate-to-strong coupling regime.

\begin{figure}[htb]
\centerline{\includegraphics[width=8.7cm]{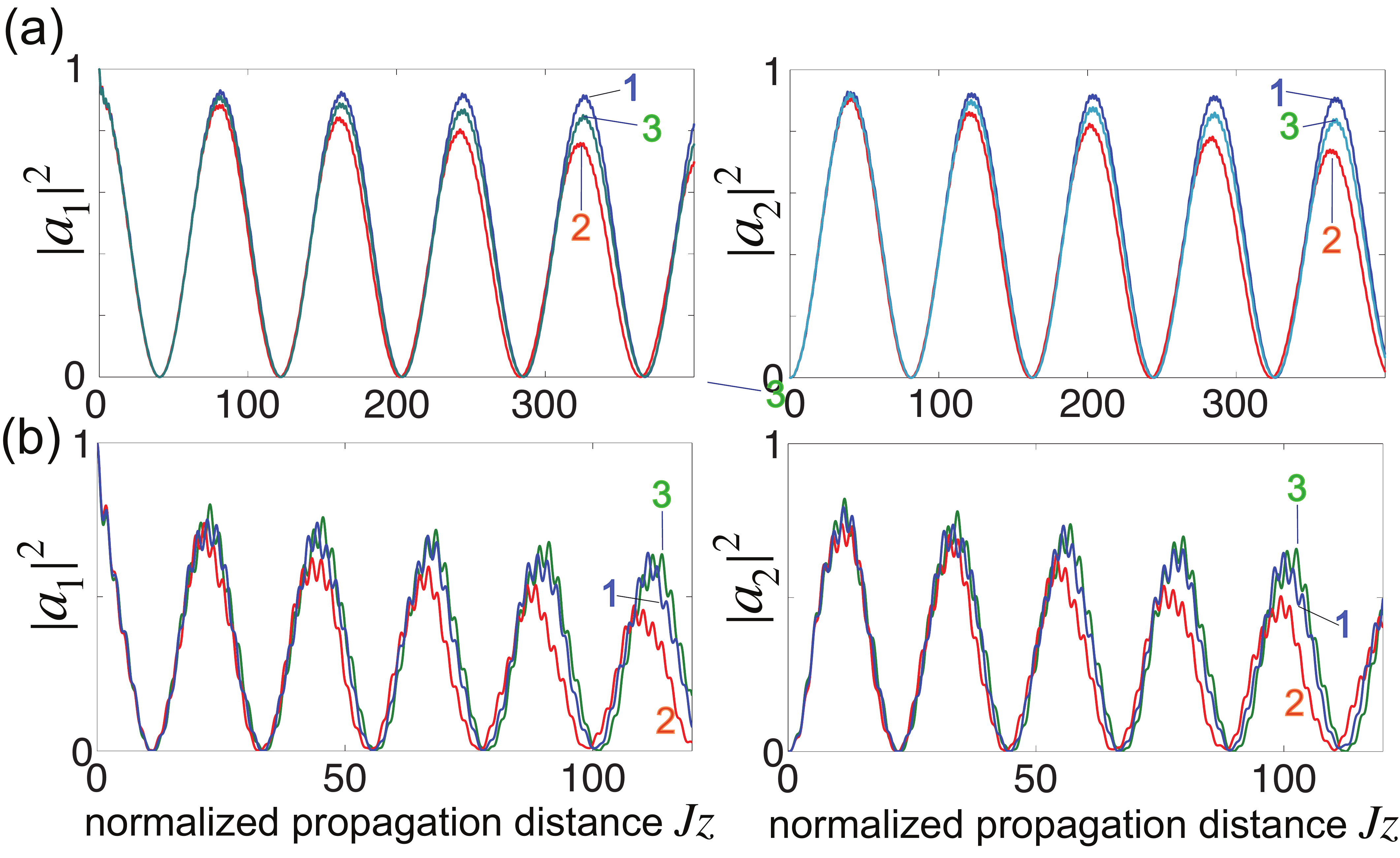}} \caption{ \small
(Color online) Decay dynamics in waveguides W$_1$ and W$_2$ (behavior of $|a_{1,2}(z)|^2$ versus normalized propagation distance $Jz$) in the waveguide-lattice system of Fig.1(e) for $h=1$ and for
(a) $\rho/J=0.2$, and (b) $\rho/J=0.4$. The different curves (labelled by 1,2 and 3) refer to different values of the propagation constant shift $\omega_0$. Curve 1: $\omega_0=-J$; curve 2: $\omega_0=-1.1 J$; curve 3: $\omega_0=-0.9 J$.}
\end{figure} 

The Rabi flopping dynamics should be feasible for an experimental observation using optical waveguide lattices realized by the femtosecond (fs) laser writing technology \cite{r7,r8,r11,r12}. For example, let us assume passive optical waveguides manufactured in fused silica and probed in the red ($\lambda=633$ nm). We numerically simulated light propagation in the waveguide lattice by solving the three-dimensional optical Schr\"odinger equation \cite{r6} using a standard pseudo spectral split-step method. In the simulations, we assumed a circular profile of the guide core with a super-Gaussian profile of radius $2 \; \mu m$, a peak refractive index change $\Delta n=1.5 \times 10^{-3}$ for the guides in the array, and a substrate refractive index $n_s=1.5$. A design as in Fig.1(e) has been assumed, with uniform spacing $a=12 \; \mu$m between adjacent guides in the array and a spacing $d=15 \; \mu$m between guide W$_{1,2}$ and the three closest guides in the array [Fig.3(a)]. This corresponds to coupling constants $J \simeq 2.15 \; {\rm cm}^{-1}$ and $\rho \simeq 1.05 \; {\rm cm}^{-1}$. The condition $\omega_0=-J$ is obtained by slightly decreasing the peak refractive index change for the guides W$_{1,2}$ to the value $\Delta n_1=1.4285 \times 10^{-3}$. Figure 3(b) shows the numerically-computed fractional optical power trapped in the two guides  W$_{1,2}$ as a function of the propagation distance, clearly showing the appearance of Rabi oscillations, which a period of about $\sim 7$ cm. In an experiment using fs-laser-written arrays, some difficulties could arise from fine tuning of coupling constant $\rho$ (mainly because of coupling anisotropy arising from the non-circular shape of the guide core) and of the refractive index change in guides W$_{1,2}$, which is usually achieved by  velocity control of the fs laser writing beam. However, as shown in Fig.2(b) even for moderate deviations from the resonance condition the lifetime of the quasi BICs remain
long enough to observe several cycles of Rabi flopping.

\begin{figure}[htb]
\centerline{\includegraphics[width=8.7cm]{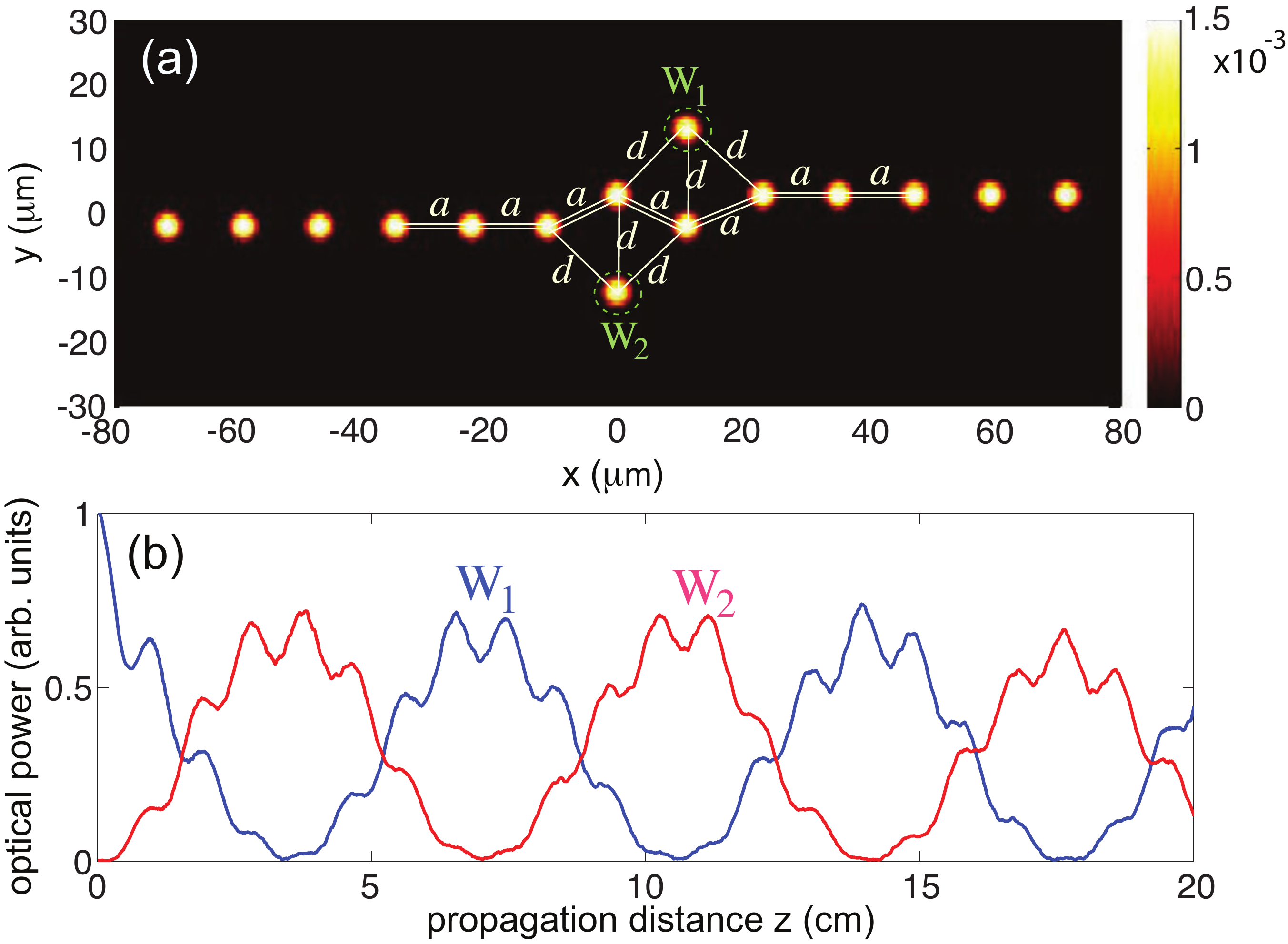}} \caption{ \small
(Color online) (a) Refractive index profile $n(x,y)-n_s$ of the waveguide-array system in the transverse $(x,y)$ plane, on a pseudo color map, used in numerical simulations. (b) Numerically-computed behavior of the fractional optical power trapped in waveguides W$_1$ and W$_2$ versus propagation distance $z$. The system is excited at $z=0$ in the fundamental mode of waveguide W$_1$ at $\lambda=633$ nm.}
\end{figure} 
{\em Conclusions.} We introduced the idea of Rabi flopping between BICs, indirectly coupled via a continuum, and proposed an integrated optics setup where such effect could be observed. The present results unravel new insights into BIC coupling and could be of interest beyond photonics, for example in waveguide quantum electrodynamics with superconductiong circuits \cite{r36}, where BIC coupling could provide a decoherence-free subspace.\\
\\

\end{document}